\documentclass[12pt,epsf]{article}

\usepackage{graphicx}

\begin{document}
\title{$b \to ss\bar{d}$ in a Vector Quark Model}
\author{Hai-Ying Cai\footnote{hycai@gms.phy.pku.edu.cn}
and Da-Xin Zhang\footnote{dxzhang@mail.phy.pku.edu.cn}\\
\it\small Institute of Theoretical Physics,\\[-3mm]
\it\small School of Physics, Peking University, Beijing 100871,
China}
\date{February 6, 2004}

\maketitle

\begin{abstract}
The rare decay $b\to ss\bar{d}$ is studied in a vector quark model
by adding the contributions from exotic vector-like quarks. We
find that the contribution from box diagrams amounts to $10^{-9}$
in the branching ratio, while the $Z$-mediated tree level
contribution is negligible.
\end{abstract}

\vspace{0.5cm} {\it PACS}: 12.15.Mm, 13.90.+i, 14.40.Lb, 14.40.Nd

\newpage

\section{Introduction}
The flavor changing neutral current (FCNC) processes in B physics
provide  important windows to expose potential signals induced by
new physics up to a scale around TeV. In the standard model (SM),
the FCNC processes are induced at loop levels and are further
suppressed by the Glashow-Maiani-Illiopoulous (GIM) mechanism. It
is possible for the contribution from the new physics, whether at
tree level or at loop level,  to be competitive to its
corresponding SM backgrounds. We could find out possible new
physics through searching for the deviations from the SM
predictions. It is not surprising that various FCNC processes are
extensively studied in many of the extended models beyond the SM.
Suppressed strongly in the SM as a second order weak process with
strong GIM cancellations, the rare decay $b \to ss\bar{d}$\cite{1}
need to be considered as an important process which provides
possible virtual signals of new physics. Many new physics models
\cite{2} with different flavor structures have shown the potential
of enhancing significantly the branch ration of $b \to ss\bar{d}$.
The corresponding exclusive decays ($e.g.$ $B^-\to K^-K^-\pi^+$)
have also been searched experimentally by different groups
\cite{data}, which provides further constrains on the new physics
models.

In this Letter, we investigate $b \to ss\bar{d}$ in a Vector Quark
Model (VQM)\cite{vqm}. With the inclusion of exotic heavy quarks
with different quantum numbers under the SM gauge groups, it could
be possible that the CKM matrix elements in the VQM  are different
and that the GIM cancellations in the first three generations are
violated by these extra heavy quarks. Demanding all existed
phenomenological constrains satisfied, we find  that the branch
ration of $b \to ss\bar{d}$ in a VQM could  amount to $10^{-9}$,
several orders larger than its SM prediction which is below
$10^{-12}$\cite{1}.

\section{Brief Review of Vector Quark Model}

The VQMs are  the SM extensions  by adding into exotic quarks with
non-standard $SU(2)_L\times SU(1)_Y$ assignments. The models could
naturally emerge from  some extensions of SM such as $E_6$ grand
unified theory. Although these exotic quarks are heavy, they do
not necessarily  decouple in the low energy phenomenology. At low
energy they exhibit their effects through mixing with the ordinary
quarks of the first three generations.

Here we focus on  a simple model with  one extra $Q=\frac{2}{3}$
up-type vector-like quark and one extra $Q=-\frac{1}{3}$ down-type
vector-like quark, both of their left-hand and right-hand
components translate as singlets under the SM gauge group.
The ordinary  and the vector-like quarks of the same electrical
charges mix into the mass eigenstates which are denoted as
\begin{eqnarray}
(u_{L,R})_\alpha=\left[
\begin{array}{c}u_{L,R}\\c_{L,R}\\t_{L,R}\\T_{L,R}\end{array}
\right]_\alpha, ~~(d_{L,R})_\alpha=\left[
\begin{array}{c}d_{L,R}\\s_{L,R}\\b_{L,R}\\B_{L,R}\end{array}
\right]_\alpha, ~~(\alpha=1,2,3,4).
\end{eqnarray}
They are related to the ordinary quarks in the weak eigenstates
$u_{L,R}^0$ and $d_{L,R}^0$ by
\begin{eqnarray}
(u^0_{L,R})_i=(U^{u}_{L,R})_{i\alpha}(u_{L,R})_\alpha, ~~
(d^0_{L,R})_i=(U^{d}_{L,R})_{i\alpha}(d_{L,R})_\alpha,
~~(i=1,2,3),
\end{eqnarray}
where $U^{u}_{L,R}$ and $U^{d}_{L,R}$ are all $3\times 4$
matrices.

The charged current interactions  in the mass eigenstates are:
\begin{eqnarray}
{\cal L}_{W}=\frac{g}{2\sqrt{2}}\overline{u}_{L\alpha}\gamma^{\mu}
V_{\alpha\beta}d_{L\beta}W_{\mu}^{\dagger} +h.c.,
\end{eqnarray}
and
\begin{equation}
{\cal L}_{G^{\pm}} =\frac{g}{\sqrt{2}}\overline{u}_{\alpha}
V_{\alpha\beta}\left[\frac{m_{u\alpha}}{M_W}P_L-\frac{m_{d\beta}}{M_W}P_R\right]
d_{\beta}G^{\dagger}+h.c.
\end{equation}
in the $R_\xi$ gauge, where
\begin{eqnarray}
 V_{\alpha\beta}\equiv (U^{u\dagger}_L)_{\alpha
i}(U^d_L)_{i\beta}, ~~(i=1,2,3, \alpha, \beta=1,2,3,4)
\end{eqnarray}
is called the extended CKM matrix which is no longer unitary,
\begin{eqnarray}
(V^{\dagger}V)_{\alpha\beta}&=&
\delta_{\alpha\beta}-(U^{d\dagger}_L)_{\alpha
4}(U^d_L)_{4\beta},\nonumber \\
(VV^{\dagger})_{\alpha\beta}&=&
\delta_{\alpha\beta}-(U^u_L)_{\alpha 4}(U^{u\dagger}_L)_{4\beta}.
\label{vv}
\end{eqnarray}
The neutral current interactions are modified to
\begin{equation}
{\cal L}_{{\cal
Z}}=\frac{g}{2C_{W}}(\overline{u}_{L\alpha}\gamma^{\mu}
X^u_{\alpha\beta}u_{L\beta}-\overline{d}_{L\alpha}\gamma^{\mu}
X^d_{\alpha\beta}d_{L\beta} -2S_{W}^{2}{\cal J}_{EM}^{\mu}
)Z_{\mu},\label{zmed}
\end{equation}
where
\begin{eqnarray}
(X^u)_{\alpha\beta}&\equiv& (U^{u\dagger}_L)_{\alpha
i}(U^u_L)_{i\beta}=(VV^{\dagger})_{\alpha\beta},\nonumber \\
(X^d)_{\alpha\beta} &\equiv& (U^{d\dagger}_L)_{\alpha
i}(U^d_L)_{i\beta}=(V^{\dagger}V)_{\alpha\beta}. \label{X}
\end{eqnarray}
The neutral interactions mediated by the goldstone boson $G^0$ are
proportional to the quark masses. We will not display these small
effects, as for the process discussed here is concerned.
 It is clear from (\ref{vv} - \ref{X}) that $Z$-mediated FCNC
interactions are induced at tree level in the VQM. In (\ref{zmed})
the electromagnetic currents ${\cal J}_{EM}$ are the same as in
the SM  for the ordinary quarks. (See \cite{vqm} for details.)

\section{$b\to ss \bar{d}$}

In the SM, the main contribution to the process $b\to ss \bar{d}$
is from the box diagrams with $W$ and the up-quarks in the loops
\cite{1}. Due to the GIM mechanism, the amplitude is suppressed
either by a small factor $V_{ts}^{*}V_{tb}V_{ts}^{*}V_{td}$, where
$V_{ij}$'s stand for the (unitary) CKM matrix elements in the SM,
or by a small power factor $m_c^2/m_W^2$. The resulting branching
ratio is smaller than $10^{-12}$.

In the VQM,  $b\to ss \bar{d}$ can be induced by two mechanisms.
One is the $Z$-mediated tree diagram. The other is the box
diagrams with $W^\pm, G^\pm$ boson and the $u,c,t,T$ quarks inside
the loops. The $Z$-penguin diagrams are taken as higher order
corrections to the tree diagram and their effects need not to be
considered. We have
\begin{eqnarray}
\Gamma_{VQM}(b\to ss \bar{d})&=&\frac{m_b^5}{48(2\pi)^3}\bigg|
\frac{G_F}{\sqrt{2}}X_{sb}X_{ds}+\frac{G_F^2}{2\pi^2}m_W^2\bigg[X_{sb}X_{ds}\nonumber\\
&&+\sum_{\alpha=c,t,T}4X_{sb}\lambda^{\alpha}_{ds}B_0(x_\alpha)
+\sum_{\alpha=c,t,T}4X_{ds}\lambda^{\alpha}_{sb}B_0(x_\alpha)\nonumber\\
&&+\sum_{\alpha,\beta=c,t,T}\lambda^{\alpha}_{sb}\lambda^{\beta}_{ds}
S_0(x_\alpha,x_\beta)\bigg]\bigg|^2.\label{br}
\end{eqnarray}
We have denoted
\begin{eqnarray}
\lambda_{d_id_j}^\alpha=V_{\alpha d_i}^*V_{\alpha d_j},
~x_\alpha=\frac{m_\alpha^2}{m_W^2}.
\end{eqnarray}
 On the RHS of (\ref{br}), the term outside the bracket
represents the tree diagram contribution. In the bracket, the
first term origins from the box diagram with two $u$-quarks in the
loop; the second term is from the box diagram with $u$-quark
connected to the $s$ and $b$ legs and $c,t,T$ quarks connected to
the $d$ and $s$ legs, while the third term comes from the box
diagram with $u$-quark connected to the $d$ and $s$ legs and
$c,t,T$ quarks connected to $s$ and $b$ legs. The last term is
from the box diagrams without $u$-quark in the loop.  The
Inami-Lim functions are \cite{ilf}
\begin{eqnarray}
F(x_{\alpha},x_{\beta})&=&
\frac{4-7x_{\alpha}x_{\beta}}{4(1-x_{\alpha})(1-x_{\beta})}
+\frac{4-8x_{\beta}+x_{\alpha}x_{\beta}}{4(1-x_{\alpha})^{2}(x_{\alpha}-x_{\beta})}
x_{\alpha}^{2}\ln x_{\alpha}\nonumber\\
&+&\frac{4-8x_{\alpha}+x_{\alpha}x_{\beta}}{4(1-x_{\beta})^{2}(x_{\beta}-x_{\alpha})}
x_{\beta}^{2}\ln x_{\beta},
\end{eqnarray}
\begin{equation}
S_0(x_{\alpha})=F(x_{\alpha},x_{\alpha})-2F(0,x_{\alpha})+F(0,0),
\end{equation}
\begin{equation}
S_0(x_{\alpha},x_{\beta})=F(x_{\alpha},x_{\beta})-F(0,x_{\alpha})-F(0,x_{\beta})+F(0,0),
\end{equation}
\begin{equation}
4B_0(x_{\alpha})=F(0,x_{\alpha})-F(0,0).
\end{equation}

\section{Numerical Analysis}

The VQM model are mostly constrained  by $\Delta{\cal M}_K$,
$\Delta {\cal M}_{B_d}$ and ${\cal B}r(B\to X_s \gamma)$. In the
VQM, they can be expressed as:
\begin{eqnarray}
\Delta {\cal
M}_K^{VQM}&=&\frac{G_F}{3\sqrt{2}}m_K(B_KF_K^2)\bigg|\eta_Z^K
X_{ds}^2+\frac{G_F}{\sqrt{2}\pi^2}m_W^2\bigg[\eta_Z^K X_{ds}^2
\nonumber\\&&+\sum_{\alpha=c,t,T}8X_{ds}\lambda^{\alpha}_{ds}\eta_{\alpha\alpha}^KB_0(x_\alpha)
+\sum_{\alpha,\beta=c,t,T}\lambda^\alpha_{ds}\lambda^\beta_{ds}\eta_{\alpha\beta}^K
S_0(x_\alpha,x_\beta)\bigg]\bigg|,
\end{eqnarray}
\begin{eqnarray}
\Delta {\cal
M}_{B_d}^{VQM}&=&\frac{G_F}{3\sqrt{2}}m_{B_d}(B_{B_d}F_{B_d}^2)
\bigg|\eta_Z^BX_{db}^2+\frac{G_F}{\sqrt{2}\pi^2}m_W^2\bigg[\eta_Z^BX_{db}^2\nonumber\\&&+
\sum_{\alpha=t,T}8X_{db}\lambda^{\alpha}_{db}\eta_{\alpha\alpha}^B
B_0(x_\alpha)+\sum_{\alpha,\beta
=t,T}\lambda^\alpha_{db}\lambda^\beta_{db}\eta_{\alpha\beta}^B
S_0(x_\alpha,x_\beta)\bigg]\bigg|.
\end{eqnarray}
As for the rare decay  $B\to X_s \gamma$, it has been discussed in
reference \cite{7}. In the above equations, $\eta's$ are the QCD
factors. Here we take the values:
$\eta_Z^K=0.60$,$\eta_{cc}^K=1.38$,$\eta_{tt}^K=0.57$,$\eta_{ct}^K=0.47$,$\eta_{TT}^K=0.58$,
$\eta_{cT}^K=0.47$,$\eta_{tT}^K=0.58$;$\eta_Z^B=0.57$,$\eta_{tt}^B=\eta_{TT}^B=\eta_{tT}^B=0.55$\cite{qcd,6}.
Other parameters are $m_K=498MeV$, $F_K=160MeV$, $B_K=0.86$,
$m_{B_d}=5.279GeV$,$F_{B_d}\sqrt{B_{B_d}}=200MeV$\cite{9}.

It could be seen from (\ref{br}) that $\Gamma_{VQM}$ is
parameterized by $ X_{sb}$, $X_{ds}$,$\lambda^{c,t,T}_{ds}$ and
$\lambda^{c,t,T}_{sb}$. For simplicity, we  take all the
parameters as real in the numerical calculation. These parameters
are not independent and can be related by the extended CKM matrix.

Numerical analysis is done in the following way. We take the upper
sector of the extended CKM  matrix as $V_{ud}= 0.9721,V_{us}=
0.215,V_{ub}= 2\times 10^{-3},V_{cd}= 0.209,V_{cs}= 0.966,V_{cb}=
3.8\times 10^{-2}$. They are assumed to take their minimal values
indicated by \cite{9} so that the effects of the vector-like
quarks can reach their maximum values. The other parameters are
scanned in the regions of $200<m_T<800$,
$0<\left|V_{td}\right|<0.09$, $0<\left|V_{ts}\right|<0.12$,
$0.58<\left|V_{tb}\right|<0.99$,\cite{9}
$\left|X_{db}\right|<0.0011$, $\left|X_{sb}\right|<0.0011$ and
$\left|X_{ds}\right|<0.00001$\cite{6}. Regarding Equation[5], we
require $\left|0.9887+V_{td}^2+V_{Td}^2\right|<1.0$,
$\left|0.9794+V_{ts}^2+V_{Ts}^2\right|<1.0$ and
$\left|0.001448+V_{tb}^2+V_{Tb}^2\right|<1.0$, and use these
conditions to find out the ranges of $V_{Td}$,$V_{Ts}$ and
$V_{Tb}$. The experimental constraints on $0<\Delta{\cal
M}_K^{VQM}<2\times 3.491\times 10^{-15}$,
 $\left|\Delta{\cal M}_{B_d}^{VQM}-3.2\times 10^{-13}\right|<0.092\times 10^{-13}$,
$\left|{\cal B}r(B\to X_s \gamma)-3.15\times
10^{-4}\right|<0.54\times 10^{-4}$ are demanded. The allowed
parameter space is thus determined and  branch ratio of $b\to ss
\overline {d}$ is calculated.

\begin{table}[!htb]
\caption{ The maximum branch ration of $b \to ss\bar{d}$ vs
$m_T$.}
\begin{tabular}{lcccc}
$m_{T}(GeV)$ & $200$ & $400$ & $600$ & $800$
\\\hline $\Delta {\cal M}_K(GeV)$& $6.948\times 10^{-15}$ &
$6.926\times 10^{-15}$ & $6.758\times 10^{-15}$& $6.916\times
10^{-15}$
\\\hline $\Delta{\cal M}_{B_d}(GeV)$ & $3.234\times 10^{-13}$ &
$3.166\times 10^{-13}$ & $3.257\times 10^{-13}$ &$3.281\times
10^{-13}$\\\hline $Br(B\to X_s \gamma)$ & $3.667\times 10^{-4}$ &
$3.532\times 10^{-4}$ & $3.585\times 10^{-4}$ & $3.683\times
10^{-4}$\\
\\\hline$Br(b \to ss\bar{d})$ & $1.841\times 10^{-10}$ &
$7.719\times 10^{-10}$ & $1.308\times 10^{-9}$ & $1.903\times
10^{-9}$
\\\hline
\end{tabular}
\end{table}

We find that the contribution from the tree diagram amounts to
only $10^{-15}$ in the branching ratio, which is even negligible
compared to SM background. The effects of the box diagrams are the
main contributions and the diagram with two $T$ quarks dominates.
In Table 1, we give the branching ratio of $b\to ss \overline {d}$
along with $\Delta{\cal M}_{B_s}$ by taking $X_{sb}=0.0011$. We
also plot  the allowed branching ratio of $b\to ss \overline {d}$
as the function of $m_T$ in Figure 1.

In conclusion, we have calculated  the rare decay $b \to
ss\bar{d}$ in the VQM and find its branching ratio  could amount
to $10^{-9}$, about three orders of magnitude larger than its
corresponding SM value. This work is supported in part by the
National Natural Science Foundation of China (NSFC) under the
grant No. 90103014 and No. 10205001, and by the Ministry of
Education of China.

\begin{figure}
\begin{center}
\resizebox{16cm}{10cm}{\includegraphics{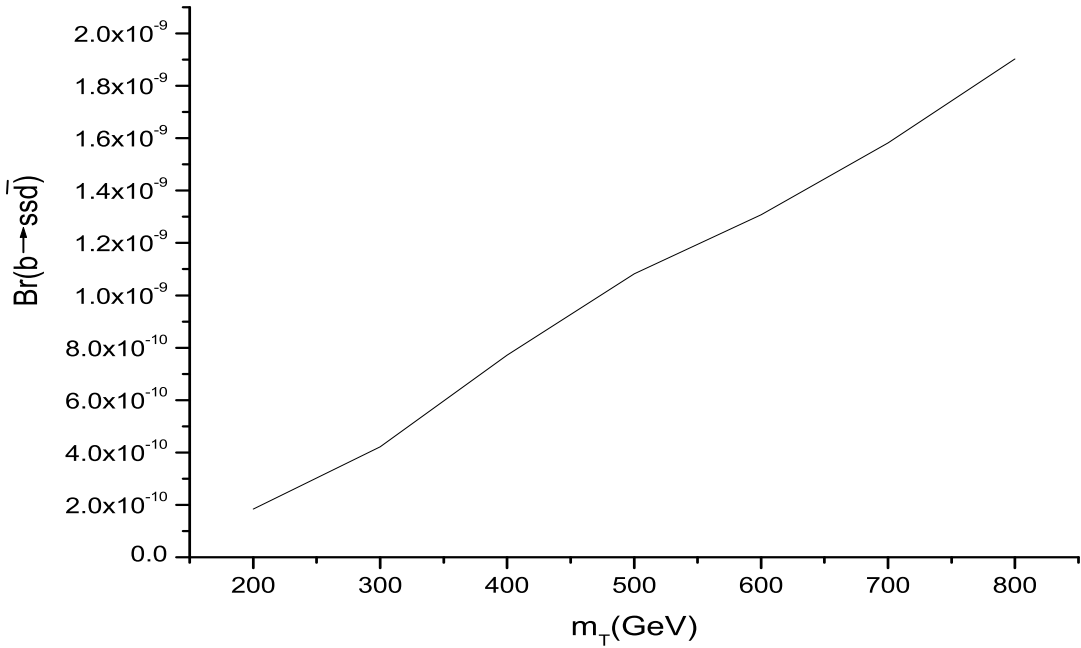}}
\end{center}
\caption{ The maximum branch ratio of $b \rightarrow s s \bar{d}$.
vs. $m_T$ }
\end{figure}

\newpage

\end{document}